\documentclass[12pt,preprint]{aastex}
\usepackage{url}
\usepackage{amssymb}
\usepackage{amsmath}
\newcommand{\be}{\begin{equation}}
\newcommand{\ee}{\end{equation}}

\shorttitle{The luminosity function during reionization} 
\shortauthors{Trenti et al.}

\begin{document}


\title{The Galaxy Luminosity Function during the Reionization Epoch}


\author{M. Trenti} \affil{CASA, Department of Astrophysics and Planetary Science, University of Colorado, 389-UCB, Boulder, CO 80309 USA} \email{trenti@colorado.edu} 

\and \author{M. Stiavelli} \affil{Space Telescope Science Institute, 3700 San Martin Drive Baltimore MD 21218 USA}
\and
\author{R.~J. Bouwens} \affil{Astronomy Department, University of California, Santa Cruz, CA 95064, USA ; Leiden Observatory, University of Leiden, Postbus 9513, 2300 RA Leiden, Netherlands}
\and
\author{P. Oesch} \affil{Institute of Astronomy, ETH Zurich, CH-8093 Zurich, Switzerland}
\and
\author{J.~M. Shull}
 \affil{CASA, Department of Astrophysics and Planetary Science, University of Colorado, 389-UCB, Boulder, CO 80309 USA}
\and
\author{G.~D. Illingworth} \affil{Astronomy Department, University of California, Santa Cruz, CA 95064, USA}
\and
\author{L.~D. Bradley}\affil{Space Telescope Science Institute, 3700 San Martin Drive Baltimore MD 21218 USA}
\and
\author{C.~M. Carollo} \affil{Institute of Astronomy, ETH Zurich, CH-8093 Zurich, Switzerland}
%

\begin{abstract}

  The new Wide Field Camera 3/IR observations on the Hubble Ultra-Deep
  Field started investigating the properties of galaxies during the
  reionization epoch. To interpret these observations, we present a
  novel approach inspired by the conditional luminosity function
  method. We calibrate our model to observations at $z=6$ and assume a
  non-evolving galaxy luminosity versus halo mass relation. We first
  compare model predictions against the luminosity function measured at $z=5$ and
  $z=4$. We then predict the luminosity function at $z\geqslant 7$
  under the sole assumption of evolution in the underlying dark-matter
  halo mass function. Our model is consistent with the
  observed $z \gtrsim 7$ galaxy number counts in the HUDF
  survey and suggests a possible steepening of the faint-end slope of the luminosity
  function: $\alpha(z \gtrsim 8) \lesssim -1.9$ compared to
  $\alpha=-1.74$ at $z=6$. Although we currently see only the
  brightest galaxies, a hidden population of lower luminosity objects
  ($L/L_{*} \gtrsim 10^{-4}$) might
 provide $\gtrsim 75\%$ of the total reionizing flux. Assuming escape
  fraction $f_{esc} \sim 0.2$, clumping factor $C\sim 5$, top
  heavy-IMF and low metallicity, galaxies below the detection limit produce complete reionization at
  $z\gtrsim 8$. 
  For solar metallicity and normal stellar IMF, reionization finishes at
  $z\gtrsim
  6$, but a smaller $C/f_{esc}$ is required for an optical depth
  consistent with the WMAP measurement.
  Our model highlights that the star formation rate in sub-$L_*$ galaxies
  has a quasi-linear relation to
  dark-matter halo mass, suggesting that radiative and mechanical feedback were less effective at $z \geq 6$ than today.

\end{abstract}


\section{Introduction}

The new Wide Field Camera 3/IR (WFC3) Hubble Ultra Deep Field (HUDF09)
observations opened a new window on high-redshift galaxy formation
\citep{oesch09_size,oesch09_zdrop,bouwens09_ydrop,bouwens09_slope,bunker09,mclure09,fink09}.
Yet the sample of $z \gtrsim 6.5$ galaxies is too small ($16$
z-dropouts in \citealt{oesch09_zdrop}, and $5$ Y-dropouts in
\citealt{bouwens09_ydrop}) for a precise determination of the galaxy
luminosity function (LF), especially after taking into account the
systematic uncertainty introduced by cosmic variance \citep{trenti08}.

Measuring the galaxy LF is important to assess their
contribution to cosmic reionization, which started at $z\gtrsim
10$, as inferred from the Thomson scattering optical depth $\tau_e$ in the CMB
background \citep{komatsu09}.
The nature of the reionizing sources is currently debated. Are normal
galaxies the agents of reionization, or are other sources responsible,
such as Population III stars or Mini-QSOs
\citep{madau04,sokasian04,shull08}? Within uncertainties, galaxies
detected at $z\sim 6$ barely keep the Universe reionized
\citep{stiavelli04b,bunker04}. The LF evolution established from
$z\sim 4$ to $z \sim 6$ \citep{bouwens07} seems to continue into the
dark ages, with progressively fewer bright sources
\citep{bolton07,bouwens08,bouwens09_ydrop,oesch09}.

The exploration of the link between LF and underlying dark-matter halo
mass function (MF) helps us understand the processes regulating star
formation. This has been studied via the conditional luminosity
function (CLF) method locally and at high redshift
\citep{vale04,cooray05a,cooray05b,cooray06,stark07,bouwens08,lee09}.
Key results are: (1) significant redshift evolution of galaxy
luminosity versus halo mass, $L(M_h)$, \citep{cooray05b,lee09}; (2)
only a fraction $\epsilon_{DC}\sim 20-30\%$ of halos appears to host
Lyman Break galaxies (LBG) \citep{stark07,lee09}; (3) the predicted LF
at $z \gtrsim 6$ deviates significantly from Schechter form, missing
the sharp drop in density of bright ($M_{AB} \lesssim -20$) galaxies
\citep{bouwens08}. These findings suggest limitations of the current
models extrapolated to the highest redshift. In fact, because of the
young age of the Universe during the reionization epoch ($\Delta z =1$
corresponds to $\lesssim 170~\mathrm{Myr}$ at $z \gtrsim 6$), it
becomes difficult to justify rapid evolution of $L(M_h)$, unless the
IMF changes. A low $\epsilon_{DC}$ also appears problematic: the halo
MF evolves rapidly at $z\gtrsim 6$: the number density of $M_h >
10^{11} M_{\sun}$ halos (hosting $\sim L_*$ galaxies) increases by a
factor three from $z=7$ to $z=6$. Hence, $\epsilon_{DC} \lesssim 0.3$
implies that the majority of recently formed halos at $z \geqslant 6$
did not experience significant star formation. Finally, the absence of
a well-defined knee in the predicted $z\gtrsim 6$ LF differs from the
rarity of observed bright galaxies (see \citealt{bouwens09_ydrop}).

To overcome these limitations, we present a novel implementation of
the CLF model, tailored for application at $z\gtrsim 5$. Instead of a
duty cycle, we adopt another simple assumption: only halos formed
within a given time interval host a detectable LBG
(Section~\ref{sec:ml}).
Section~\ref{sec:clf} contains the predictions for the $z \gtrsim 7$
LF, compared to WFC3-HUDF09 observations. Section~\ref{sec:reion}
discusses the contribution of galaxies to reionization.

\section{An Improved CLF Model for $z\sim6$}\label{sec:ml}

We adopt a variation on the CLF approach \citep{vale04,cooray05a} to
construct an empirical relation between the galaxy LF and the halo MF
at redshift $z=6$, close to the reionization epoch and with a well
measured LF function. In the standard CLF method, $L(M_h)$ is derived
assuming that each dark-matter halo hosts a single galaxy and equating
the number of galaxies with luminosity greater than $L$ to the number
of halos with mass greater than $M_h$ (optionally with $\epsilon_{DC}
\leq 1$):
\begin{equation}\label{eq:ml}
\epsilon_{DC} \int_{M_h}^{+\infty} n({M}_H,z) d{M}_H = \int_{L}^{+\infty} \phi({L},z) d{L}.
\end{equation}
Here, $n(M_h,z)$ is the \citet{st99} halo MF, constructed assuming a
WMAP5 cosmology ($\Omega_{\Lambda} = 0.72$, $\Omega_m = 0.28$,
$\Omega_b = 0.0462$, $\sigma_8=0.817$, $n_s = 0.96$, $h=0.7$;
\citealt{komatsu09}) and $\phi(L,z)$ is the galaxy LF at redshift $z$. 

\begin{figure} 
\resizebox{230pt}{!}{\includegraphics{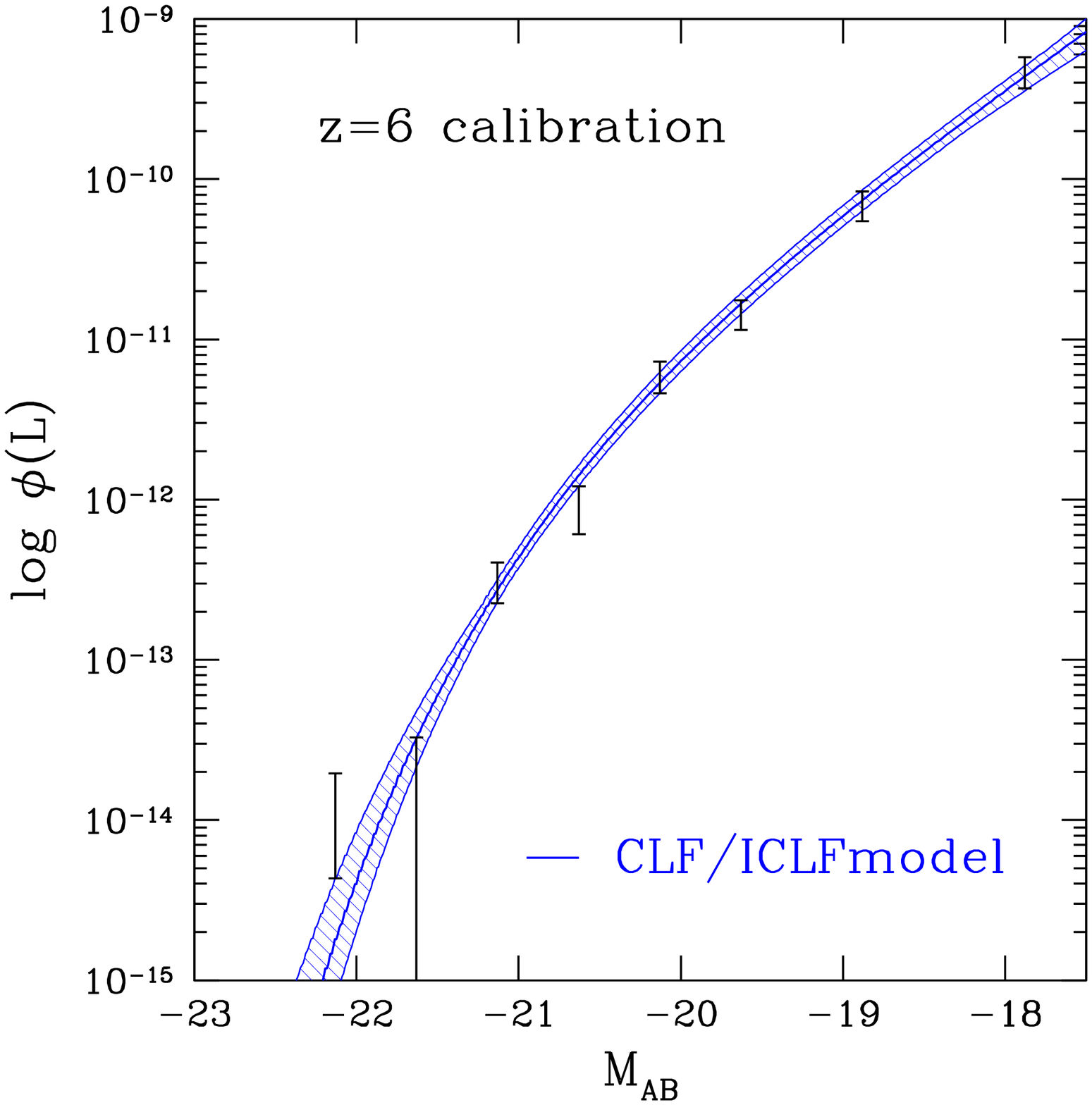}}  
\resizebox{230pt}{!}{\includegraphics{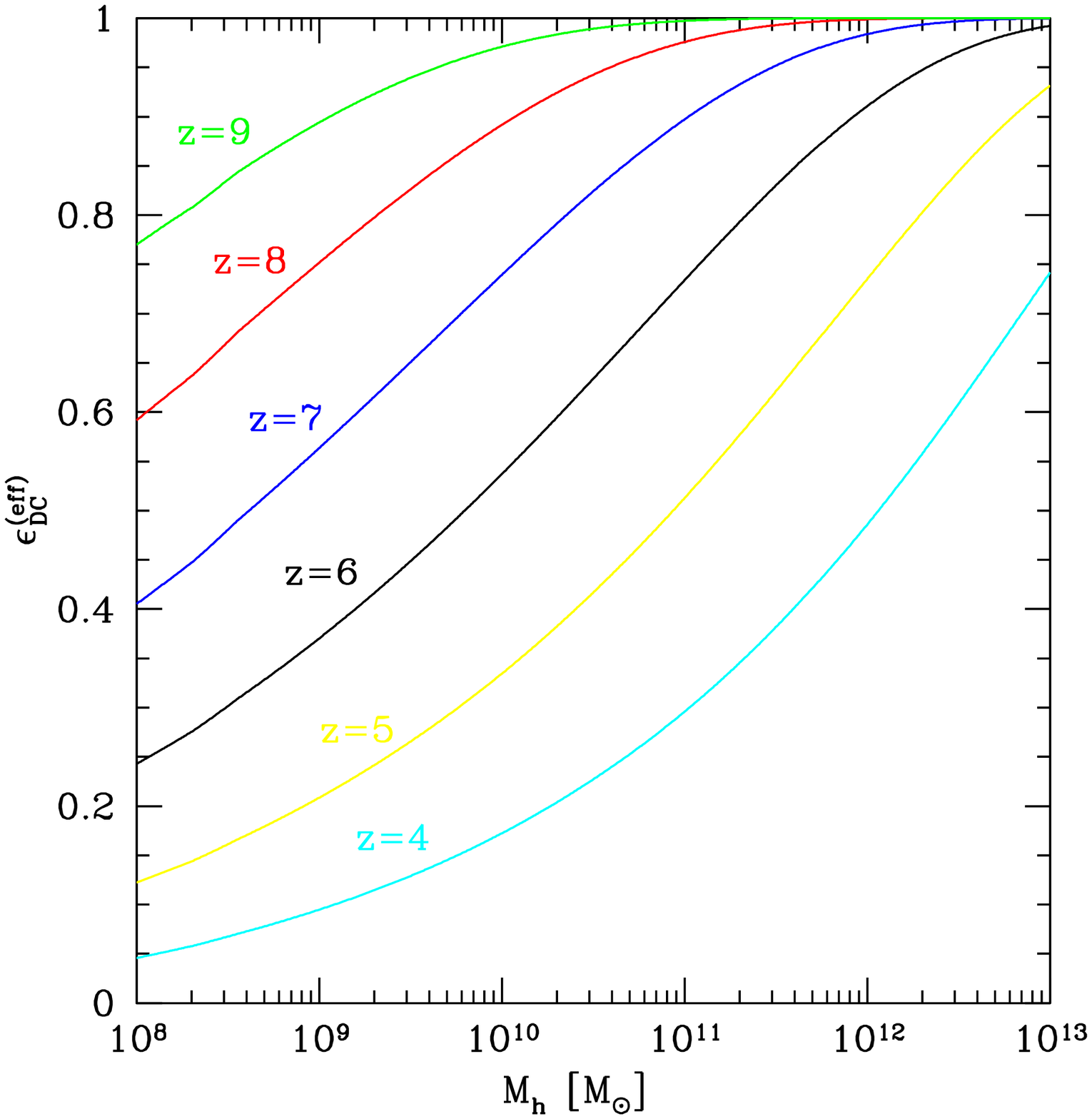}}
\resizebox{230pt}{!}{\includegraphics{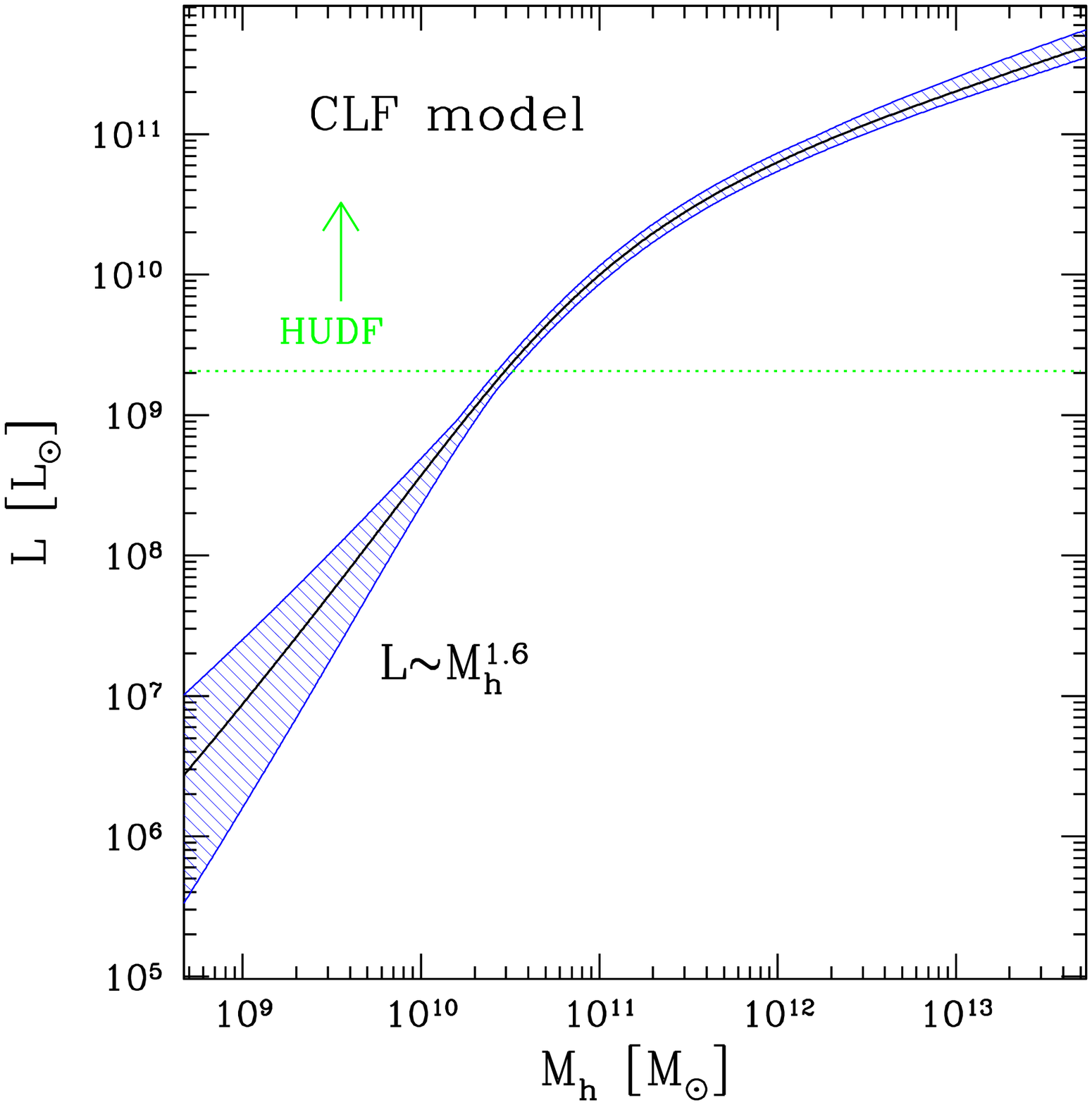}}  
\resizebox{230pt}{!}{\includegraphics{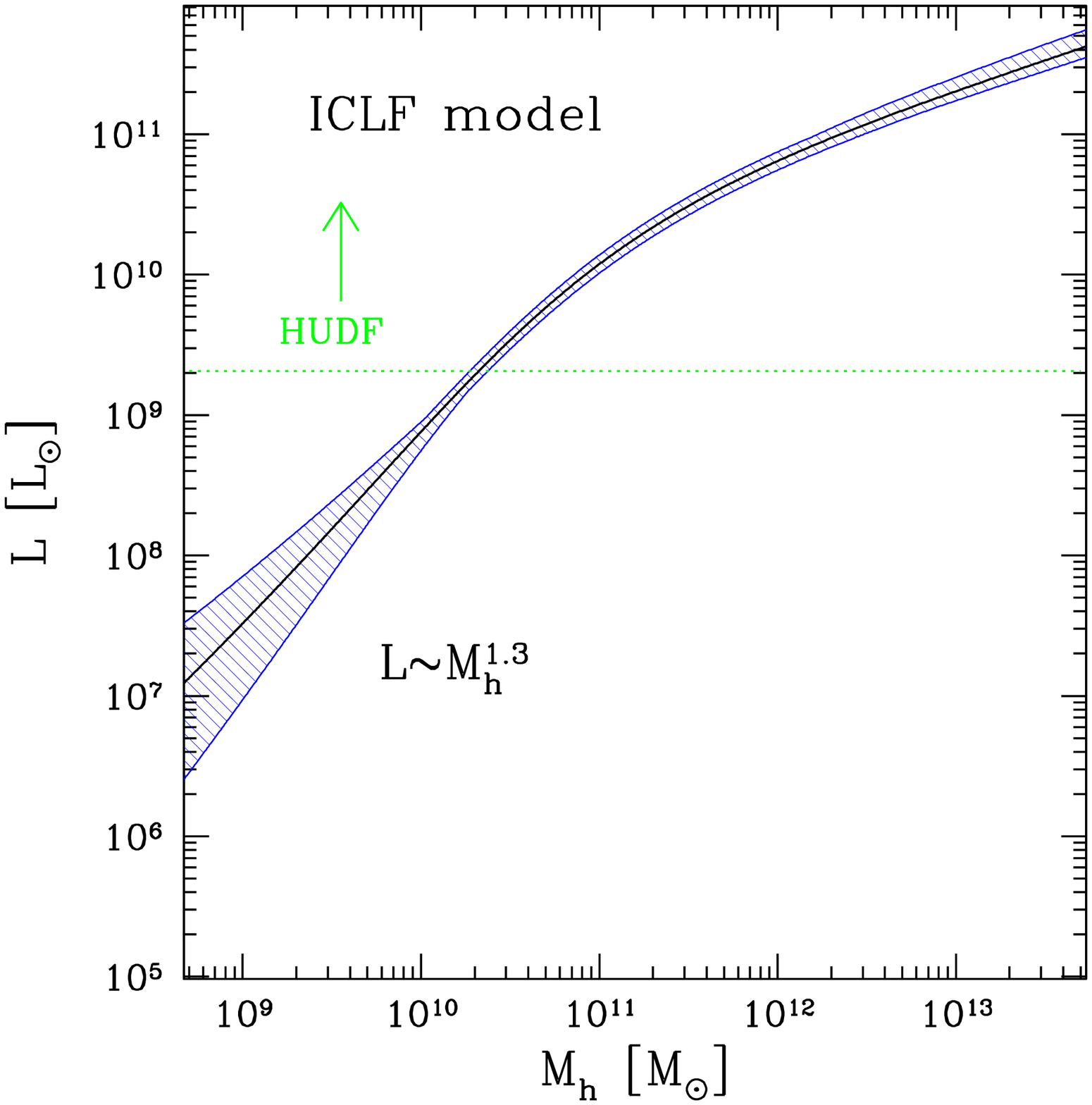}}    
\caption{Upper Right: Calibration of the CLF/ICLF models with the
  \citet{bouwens07} LF ($z=6$). Upper Left: $\epsilon_{DC}^{(eff)}(M_h,z)$ for ICLF
  model ($\Delta t =200~\rm Myr$).
 Lower panels: Galaxy luminosity versus
    dark-matter halo mass  at $z=6$ (black solid line, with blue-shaded
    area representing the 68\% confidence region). Green-dotted
 line  indicates luminosity limit $z=6$ observations in the
    HUDF. Left CLF,  right ICLF.}\label{fig:ml}
\end{figure}

In our Improved CLF (ICLF) model, rather than 
$\epsilon_{DC} < 1$, we modify Equation~\ref{eq:ml} to include only
halos with $M \geq M_h$ that have been formed within time interval
$\Delta t$:
\begin{equation}\label{eq:delta_n} \Delta N(M_h,z) =
  \int_{M_h}^{+\infty} [n({M}_H,z) - n({M}_H,z_{\Delta
    t})] \; d{M}_H, \end{equation}
where $\Delta t = t_H(z) -t_H(z_{\Delta t})$, with $t_H(z)$ being the
local Hubble time (Equation 6 of \citealt{ts09}).
Equation~\ref{eq:delta_n} defines an effective duty-cycle
$\epsilon_{DC}^{(eff)}(M_{h},z)$:
\begin{equation}\label{eq:dc_eff}
\epsilon_{DC}^{(eff)}(M_h,z) = \frac{\int_{M_h}^{+\infty} [n({M}_H,z) - n({M}_H,z_{\Delta
    t})] \; d{M}_H}{\int_{M_h}^{+\infty} n({M}_H,z) \;d{M}_H }.
\end{equation}
$L(M_h)$ is defined implicitly by:
\begin{equation}\label{eq:iclf}
\epsilon_{DC}^{(eff)}(M_h,z)  \int_{M_h}^{+\infty} n({M}_H,z) \;d{M}_H = \int_{L}^{+\infty} \phi({L},z) d{L}.
\end{equation}
In the limit $\Delta t \to +\infty$ and $\epsilon_{DC}=1$,
Equation~\ref{eq:iclf} is equivalent to Equation~\ref{eq:ml}. We adopt
$\Delta t =200~\rm Myr$ 
but discuss model predictions for $\Delta t = (100-300)~
\mathrm{Myr}$. The timescale $\Delta t$ refers to the global evolution
of the halo MF (Equations~\ref{eq:delta_n}-\ref{eq:dc_eff}) and
captures the fraction of halos that experienced a recent change in
their mass. This ensemble includes halos that have likely
experienced a recent star-formation burst, and are thus more likely to
host a UV-bright galaxy. However, for an individual halo star
formation is extended over timescales longer than $\Delta t$ at a
lower mass scale. In fact, using our Extended-Press-Schechter modeling
\citep{trenti07,tss08}, we infer that a $M_h=10^{11}~M_{\sun}$, $z=6$
halo had $M_h>10^8~M_{\sun}$, at $z > 24$ ($>99\%$ confidence). This
is consistent with the abundant supply of cold gas present at
high-redshift \citep{keres05,dave08}, which suggests sustained star
formation over several $10^8$ yr. In fact, our ICLF model has an higher
$\epsilon_{DC}^{(eff)}(M_{h},z)$ at $z\geq 6$ (Figure~\ref{fig:ml})
than the fixed $\epsilon_{DC}$ assumed/derived in similar studies
\citep{stark07,lee09}.

We parametrize the observed LF as a Schechter function:
\begin{equation}
\phi(L) = \frac{\phi_*}{L_*} \left (\frac{L}{L_*} \right)^{\alpha} \exp{(-L/L_*)}.
\end{equation}
We calibrate the model to the rest-frame UV parameters measured by
\citet{bouwens07} for i-dropouts ($z\sim 6$): $\phi_*=1.4\times
10^{-3}~\mathrm{Mpc^{-3}}$ ($h=0.7$), $\alpha=-1.74$,
$L_*=10^{-M_*/2.5}$ with $M_*=-20.24$ (see also Table~1). In
Equations~\ref{eq:ml} and \ref{eq:iclf}, we do not consider scatter in
$L(M_h)$ because we are primarily interested in the insensitive faint
end of the relation \citep{cooray05a}. We also neglect multiple halo
occupation, motivated by current $z=6$ observational limits. Only
halos with $M_h \gtrsim 7\times 10^{11}~\mathrm{M_{\sun}}$ are likely
to host multiple galaxies \citep{wechser01}. Such halos are rare
within a single ACS field of view ($\lesssim 1$ expected in the $\sim
3\times 10^4~\mathrm{Mpc^3}$ ACS volume for i-dropouts). Halos hosting
multiple galaxies are present in surveys  at $z\gtrsim6$ with a larger
volume such as the GOODS field, but their depth is insufficient to
detect the fainter (sub-$L_*$) satellite galaxies. The model of
\citet{lee09} provides an independent confirmation.

Figure~\ref{fig:ml} shows the $L(M_h)$ relation at $z=6$ from
Equations~\ref{eq:ml}-\ref{eq:iclf}. The faintest galaxies ($M_{AB}
\lesssim -18$) live in $M\gtrsim 2\times 10^{10}~\mathrm{M_{\sun}}$
halos. The blue-shaded region represents the uncertainty derived by
varying the LF parameters within the $1\sigma$ confidence regions in
Figure~3 of \citet{bouwens07}. We included an additional $12\%$
uncertainty in $\phi_*$ from  cosmic variance
($\sim 21 \% / \sqrt{3}$ for the three quasi-independent
HUDF05-ACS fields, see \citealt{trenti08}\footnote{Cosmic variance calculator available at \url{http://casa.colorado.edu/~trenti/CosmicVariance.html}}).
$L(M_h)$ is similar for both models. The steep faint-end slope of the
observed LF ($\alpha \sim -1.7$) implies a flattening of $L(M_h)$
compared to the local Universe, where $L \propto M_h^4$
\citep{cooray05a}. For the standard CLF model we derive $L \sim
M_h^{1.6}$, while the ICLF model gives $L \sim M_h^{1.3}$. This means
that the specific star formation efficiency $\eta$ in small-mass halos
depends mildly on halo mass ($\eta \sim M_h^{0.3}$), compared to the
strong suppression ($\eta \sim M_h^3$) inferred at $z=0$. This
provides clues to the processes that regulate early-time star
formation, suggesting a scenario where radiative and supernova
feedback were less efficient than today. The UV background decreases
at $z>4$ \citep{haardt96}, likely reducing the impact of
photoionization.
In addition, halos were more compact, making gas expulsion more
difficult.

From the $L(M_h)$ relation, combined with the measure of the total
stellar mass in $z=6$ galaxies based on SED fits
\citep{stark09,gonzalez09,labbe09a}, we derive a typical star
formation efficiency $\eta(M_h=10^{11} M_{\sun}) \sim 0.06$ (6\% of
gas converted into stars) with a large uncertainty, driven by the
measure of the stellar mass (Figure~9 in \citealt{stark09}).
Lower-mass halos are slightly less efficient at converting gas into
stars. For example, halos with $M_h \sim 10^8 ~M_{\sun}$, have $\eta
\sim 0.06 \times (10^8/10^{11})^{0.3} \lesssim 10^{-2}$, consistent
with assumptions in numerical models at $z\gtrsim 6$
\citep{ts09,trenti09b}. The decrease in $\eta$ with decreasing halo
mass is possibly related to suppression of star formation by local
photoionization \citep{cantalupo10}.

From $L(M_h)$ we infer a halo mass $M_h \geqslant 8\times
10^{10}~M_{\sun}$ for galaxies with $M_{AB} \leqslant -19.5$. This
agrees with clustering measurements at $z=5$ and $z=6$
\citep{overzier06}.

\subsection{Validation of the ICLF Model}

To validate the predictions of our ICLF model under the sole
assumption of evolution in the underlying dark-matter MF, we apply
$L(M_h)$ derived at $z=6$ to $\Delta N(M_h,z)$ at $z=4$ and $z=5$. The
resulting LFs are reported in Table~\ref{tab:lf} and
Figure~\ref{fig:lf}, and compared to those obtained with the standard
CLF method. As expected from past studies \citep{cooray05b,lee09}, the
standard CLF model fails to match the observed LF at $z \leq 5$ primarily
because of strong evolution in $\phi_*$ [$\phi_*(z=4) / \phi_*(z=6)
\sim 2.4$]. The ICLF model predicts instead quasi-constant $\phi_*$,
because the comoving formation rate per unit time for halos hosting
faint ($L\lesssim L_*$) galaxies remains approximately constant
between $z=6$ and $z=4$. The ICLF results are fully consistent with
the observed $z=5$ LF, and with the bright-end at $z=4$. However, the
faint-end slope at $z=4$ is underestimated at $\sim 3 \sigma$
(predicted $\alpha \gtrsim -1.6$ versus $\alpha=-1.73 \pm 0.05$
measured). This clearly indicates that our simple assumption $\Delta t
= 200 ~\mathrm{Myr}$ is no longer valid at $z\lesssim 4.5$ because
$\epsilon_{DC}^{eff}$ becomes too small at the faint end
(Figure~\ref{fig:ml}).

\section{Luminosity Function Evolution and HUDF09
  Dropouts}\label{sec:clf}

The current sample of $z\gtrsim6.5$ galaxy candidates is too small to
provide an independent fit of the LF. For example,
\citet{oesch09_zdrop} measured the $z\sim7$ faint-end slope 
assuming a fixed value for $\phi_*$ and $M_*$. Here, we apply our ICLF
model for a full prediction of the $z \geqslant 7$ LF. We do not allow
evolution of $L(M_h)$. Significant redshift evolution is present
(Figure~\ref{fig:lf}), simply because there is progressively less
structure at higher $z$. The decrease in $\phi_*$ is smaller than
predicted by the CLF method. Both models predict a dimming in $M_*$
($\partial M_* /\partial z \sim 0.25$) and a steepening of the
faint-end slope, which becomes close to $\alpha = -2$ by
$z=9$. The LF evolution is directly related to evolution of the halo
MF shape, which depends exponentially on $M_h$. As
redshift increases, massive halos become rarer, and the relative
abundance of smaller mass halos increases.

To have an accurate comparison between predicted and observed number
counts, we convolve the LF with the effective volume of the
observations as measured with artificial source recovery simulations
\citep{oesch09_zdrop,bouwens09_ydrop}. This is crucial because of
significant incompleteness at $-19 \lesssim M_{AB} \lesssim -18$.
These numbers are reported in Table~\ref{tab:hudfcounts} and include
two additional ICLF models with $\Delta t =100~\mathrm{Myr}$ and
$\Delta t = 300~\mathrm{Myr}$. The counts expected from our reference
ICLF model agree remarkably well with the number of sources observed in the
HUDF: $13.4 \pm 5.8$ z-dropouts are predicted (with $1\sigma$
uncertainty including cosmic variance) in agreement with the $16$
candidates of \citet{oesch09_zdrop}. The ICLF models with $\Delta t =
100-300~\mathrm{Myr}$ are also consistent with the data at $1\sigma$.
No LF evolution from $z=6$ predicts 31 sources (rejected at $\sim
95\%$ confidence level), while the standard CLF model gives 9.8
sources (rejected at $\sim 90\%$ confidence). For Y-dropouts, our
reference ICLF model gives $5.3$ sources at $z\sim 8$ (compared to
$24$ without LF evolution), again fully consistent within
$1\sigma$ with the $5$ Y-dropout candidates of \citet{bouwens09_ydrop}.

\section{Consequences for Reionization}\label{sec:reion}

With our LF model we can investigate the role of galaxies in the
reionization of the Universe. Despite large differences in the
estimate of reionizing photon production, past studies established
that sources below the current detection limit likely provide a
significant fraction of the photon budget. The precise number of
ionizing photons thus depends on the minimum luminosity chosen for the
extrapolation of the LF. Our modeling offers a physically motivated
cut-off for the luminosity of the smallest halo capable of forming
stars. Theoretical and numerical investigations establish that a halo
at $z\lesssim 10$ irradiated by a UV field comparable to the one
required for reionization needs a mass $M_h \gtrsim (0.6-1.7)\times 10^8
M_{\sun}$ (virial temperature $T_{vir}\gtrsim (1-2) \times 10^4$ K at
$z=7$) in order to cool and form stars \citep{tegmark97}. For $T_{vir}
\gtrsim 2\times 10^4$ K, the minimum halo mass corresponds to $M_{AB}
\approx -10$ based on the $L(M_h)$ relation at $z=6$ . Galaxies below
the HUDF09 magnitude limit $M_{AB} \sim -18$ contribute $\gtrsim 75\%$
of the total luminosity density at $z=7$ integrated to $M_{AB}=-10$,
unless feedback stronger than seen in cosmological simulations
\citep[e.g.,][]{ricotti08} induces a flattening of the LF below the
HUDF09 detection limit.

To evaluate the likelihood that galaxies ionize the universe, we
resort to the widely used conversion of luminosity density in star formation rate 
(SFR; Equation 2 in \citealt{madau98}). We compare this SFR
(Figure~\ref{fig:sfr}) to the critical rate \citep{madau99} required for reionization,
\begin{equation}\label{eq:sfr}
(SFR)_{crit} \sim 0.01 \left (\frac{0.5}{f_{esc}} \right ) \left (\frac{C}{5} \right) \left (\frac{1+z}{8}\right)^3~ M_{\sun}~\mathrm{yr^{-1}~Mpc^{-3}},
\end{equation}
where $f_{esc}$ is the escape fraction of ionizing photons and $C$ the
hydrogen clumping factor. Both Equation 2 in \citet{madau98} and
Equation~\ref{eq:sfr} above depend on explicit assumptions on the
stellar IMF (\citealt{salpeter} in $[0.1:100] M_{\sun}$ ) and
metallicity (Solar). Equation~\ref{eq:sfr} has a large uncertainty,
including the IMF-dependent efficiency of Lyman-Continuum photon
production. We adopt $f_{esc} \gtrsim 0.2$, consistent with the very
blue UV slope of small $z=7$ galaxies \citep{bouwens09_slope}, and $C
\sim 5$ \citep{bolton07,pawlik09}. Figure~\ref{fig:sfr} shows that
$z\lesssim 7$ galaxies with $M_{AB} \leqslant-10$ produce enough
photons to reionize the Universe.

To obtain the evolution of the reionization fraction, $\xi(z)$, we
follow a complementary approach (see Equation~9 and its derivation in
\citealt{stiavelli04a}). We adopt an IGM temperature $T=2 \times
10^4~\mathrm{K}$, include the effect of Helium ($\mathrm{He/H}=0.083$)
and assume conservatively $f_{esc}=0.2$, $C=5$, and no ionizing flux
at $z\geqslant 9.5$. The production rate of ionizing photons is
obtained from the LF by assuming different SEDs for the sources.
Figure~\ref{fig:sfr} shows that if the LF is integrated to
$M_{AB}=-10$, reionization by $z \sim 6$ is achieved for any SED,
including the unlikely scenario with Salpeter IMF and $Z=Z_{\sun}$
(consistent with the $(SFR)_{crit}$ analysis). However, in this case,
the optical depth to reionization is underestimated compared to the
WMAP5 measurement because of the sharp drop of $\xi(z)$ at $z\gtrsim
7$ ($\tau_e \sim 0.05$ versus $(\tau_{e})_{WMAP} = 0.084 \pm 0.016$
[\citealt{komatsu09}]). Metal-poor and top-heavy SEDs alleviate this
problem, as they achieve complete reionization at $z \gtrsim 8$,
predicting $\tau_e \gtrsim 0.07$. Alternatively, $C/f_{esc} \lesssim
10$ is needed. Without a steeper faint-end, $\alpha(z\geq 6) =
-1.74$, the $z=8$ ionizing flux is reduced by $\sim 40\%$, requiring a
corresponding decrease in $C/f_{esc}$ for a constant $\tau_e$. Without
the contribution from sources below the HUDF detection limit (i.e., LF
integration to $M_{AB}=-18$), only a top-heavy and metal-poor SED can
reionize the universe by $z=6$, but that model predicts $\tau_e\sim
0.05$.

\section{Conclusions}\label{sec:conc}

\begin{figure} 
\resizebox{180pt}{!}{\includegraphics{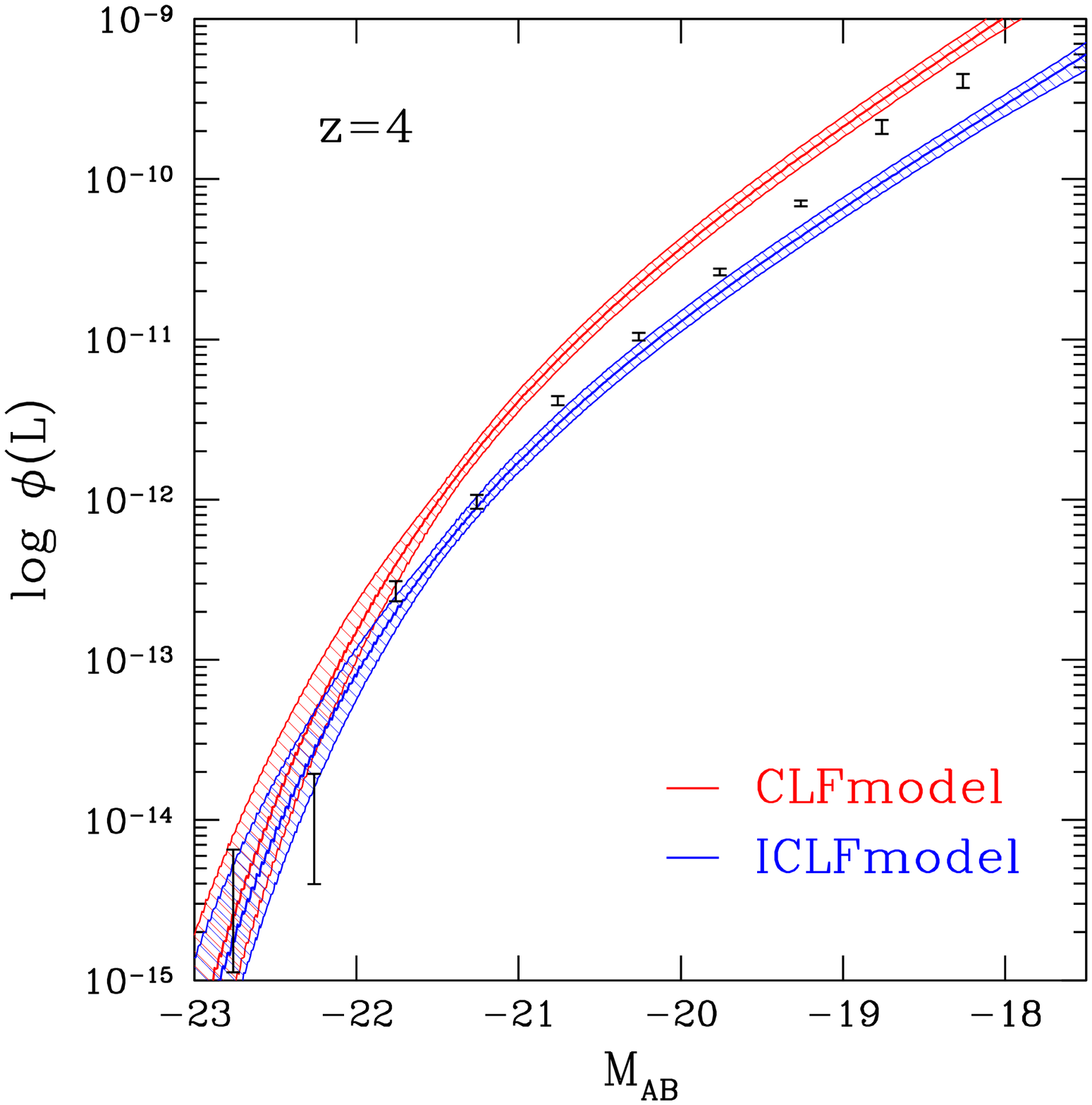}}  
\resizebox{180pt}{!}{\includegraphics{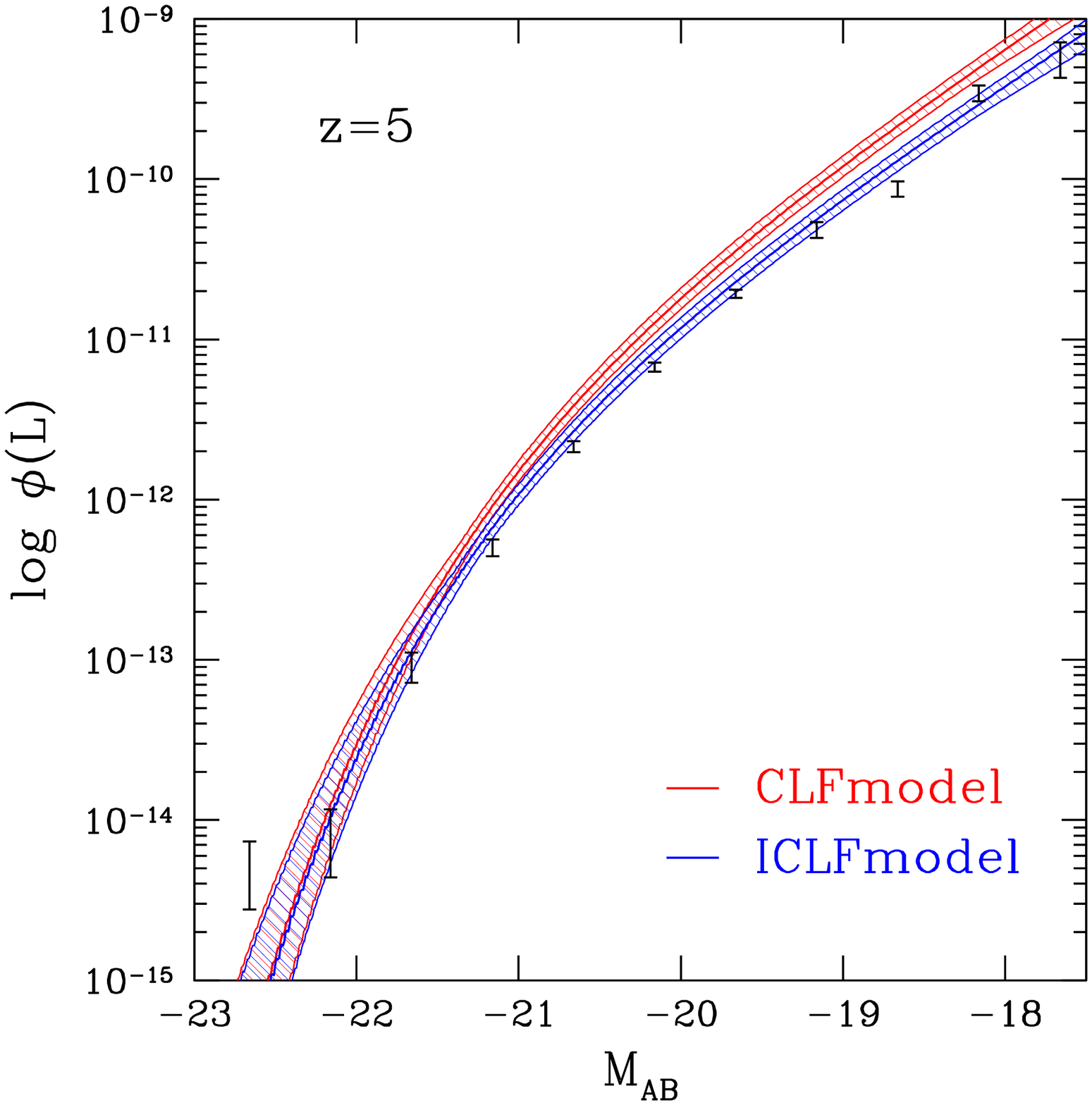}}  
\resizebox{180pt}{!}{\includegraphics{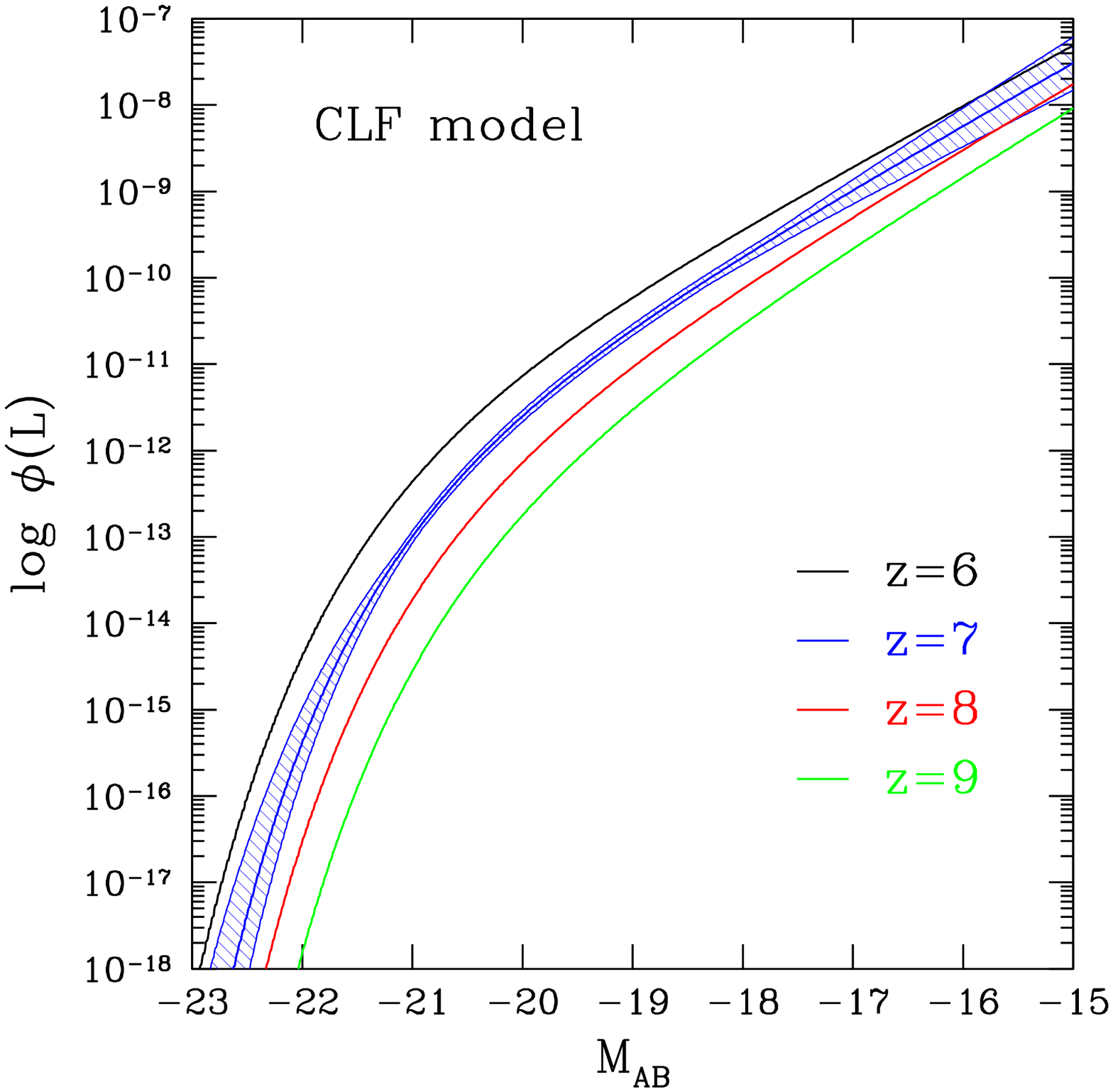}}  
\resizebox{180pt}{!}{\includegraphics{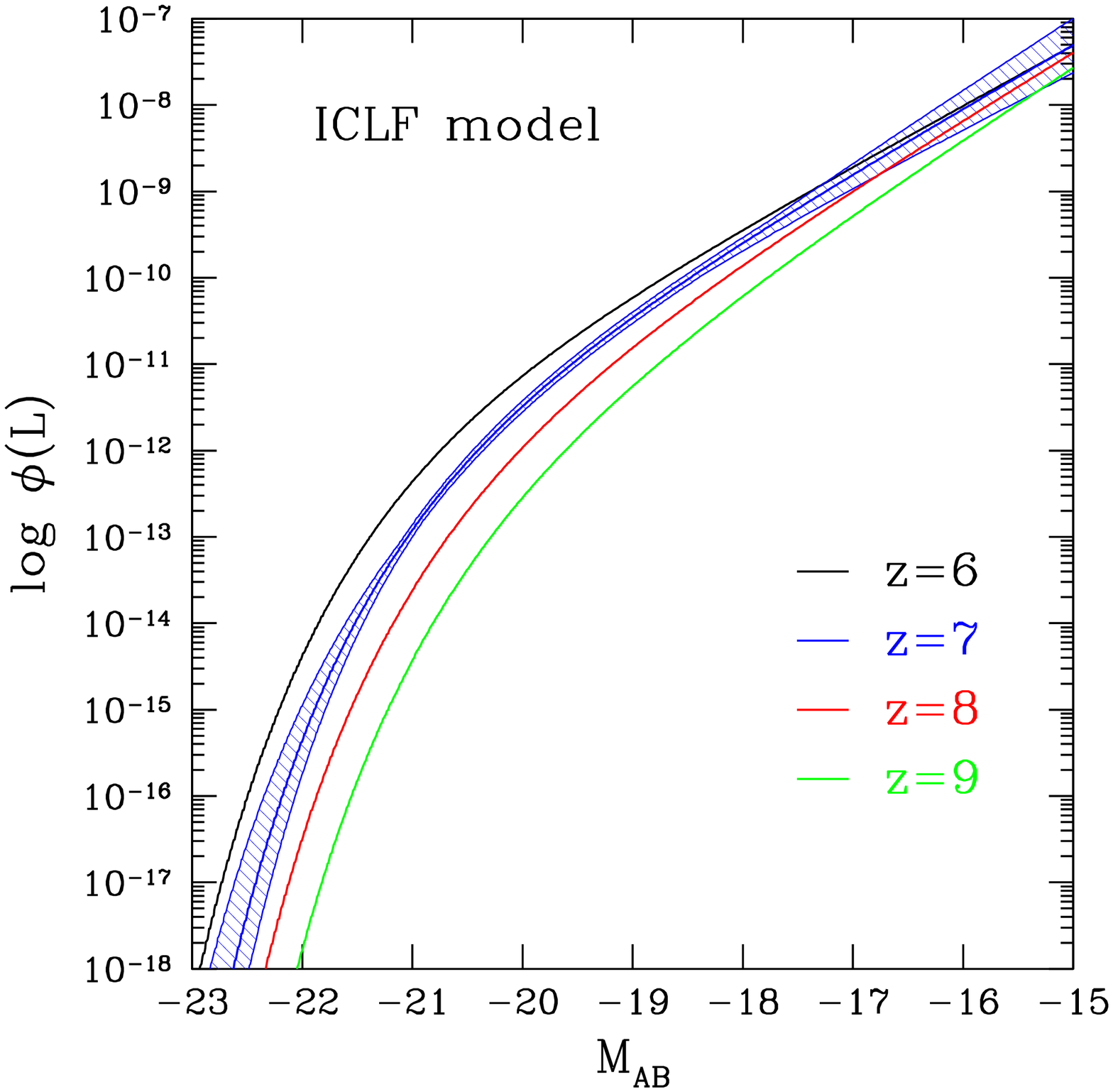}}  
\resizebox{180pt}{!}{\includegraphics{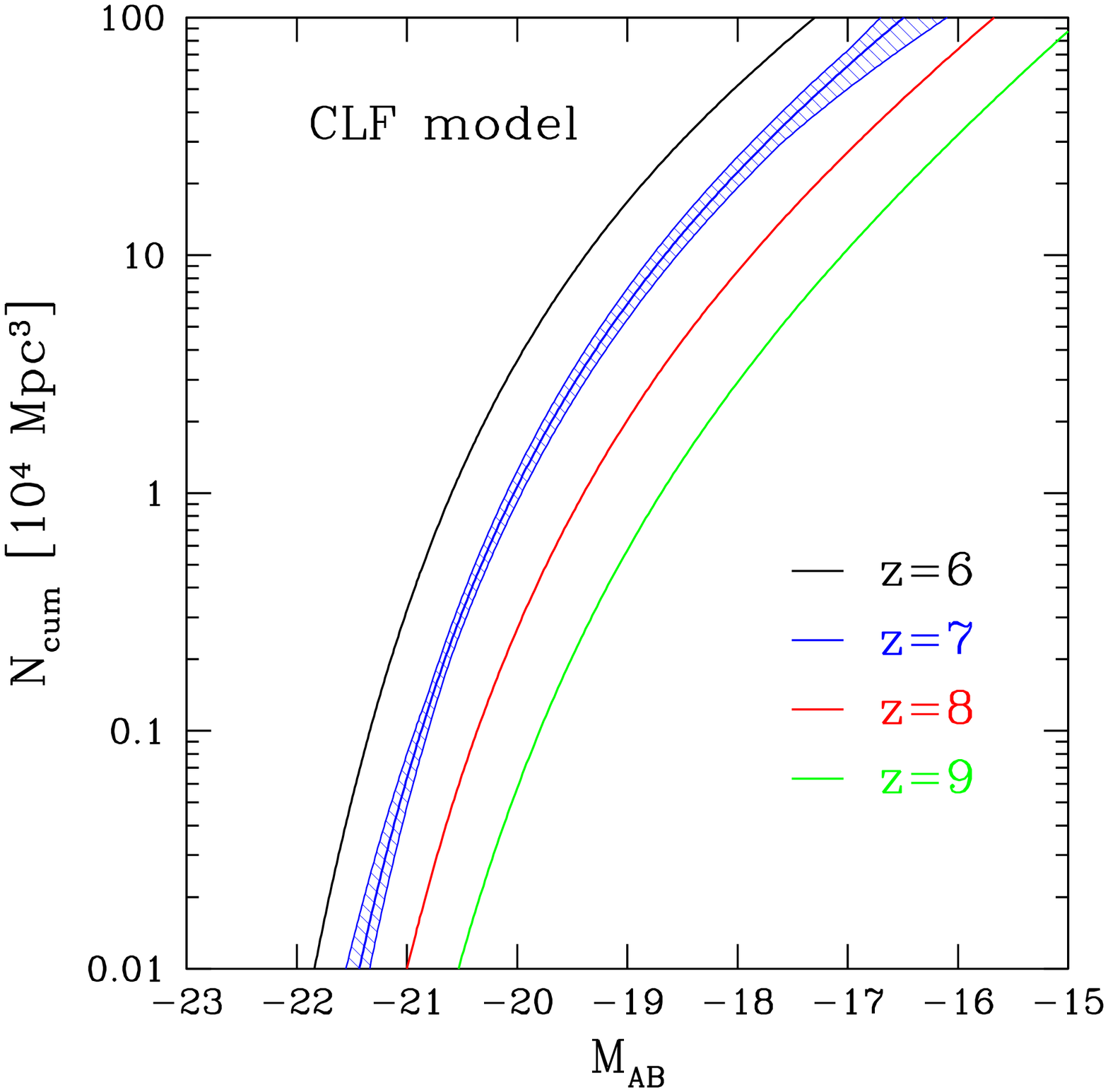}}  
\resizebox{100pt}{!}{\includegraphics{empty.eps}}  
\resizebox{180pt}{!}{\includegraphics{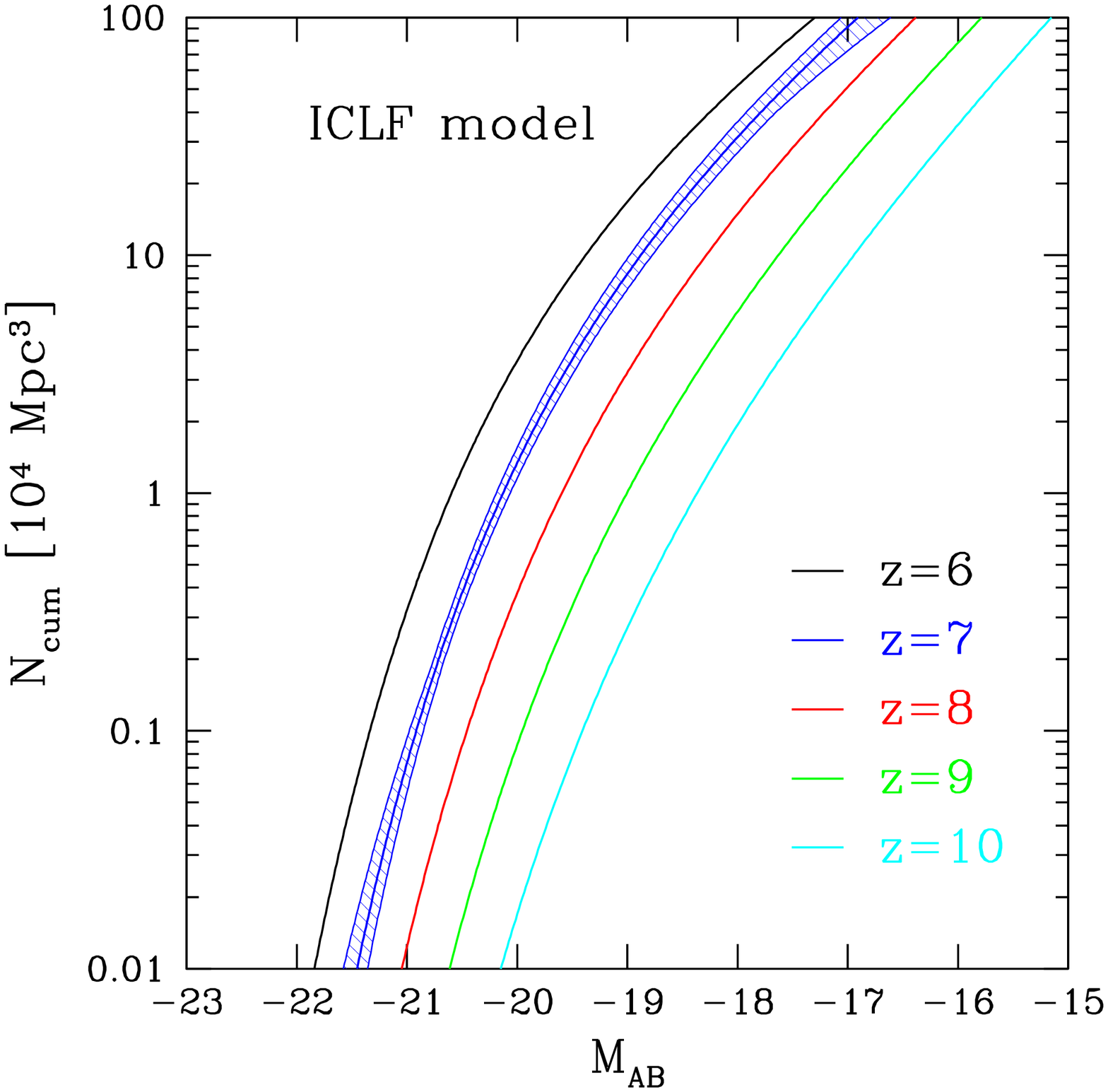}}  
\caption{Upper panels: Model comparison (red CLF, blue ICLF) at $z=4~\mathrm{(left)}$ and $z=5~\mathrm{(right)}$ with \citet{bouwens07} LF (black points).  Central panels: LF (black: $z=6$; blue $z=7$;
    red $z=8$; green $z=9$) obtained with CLF method (left) and
    our ICLF model (right). Lower panels: cumulative number of galaxies for a 
comoving volume of $10^4~\mathrm{Mpc^3}$ ($h=0.7$), similar to HUDF09 field.
  Blue-shaded area gives $1\sigma$ confidence region
   (shown for $z=7$ only).}\label{fig:lf}\label{fig:n_cum}
\end{figure}

\begin{figure} 
  \plottwo{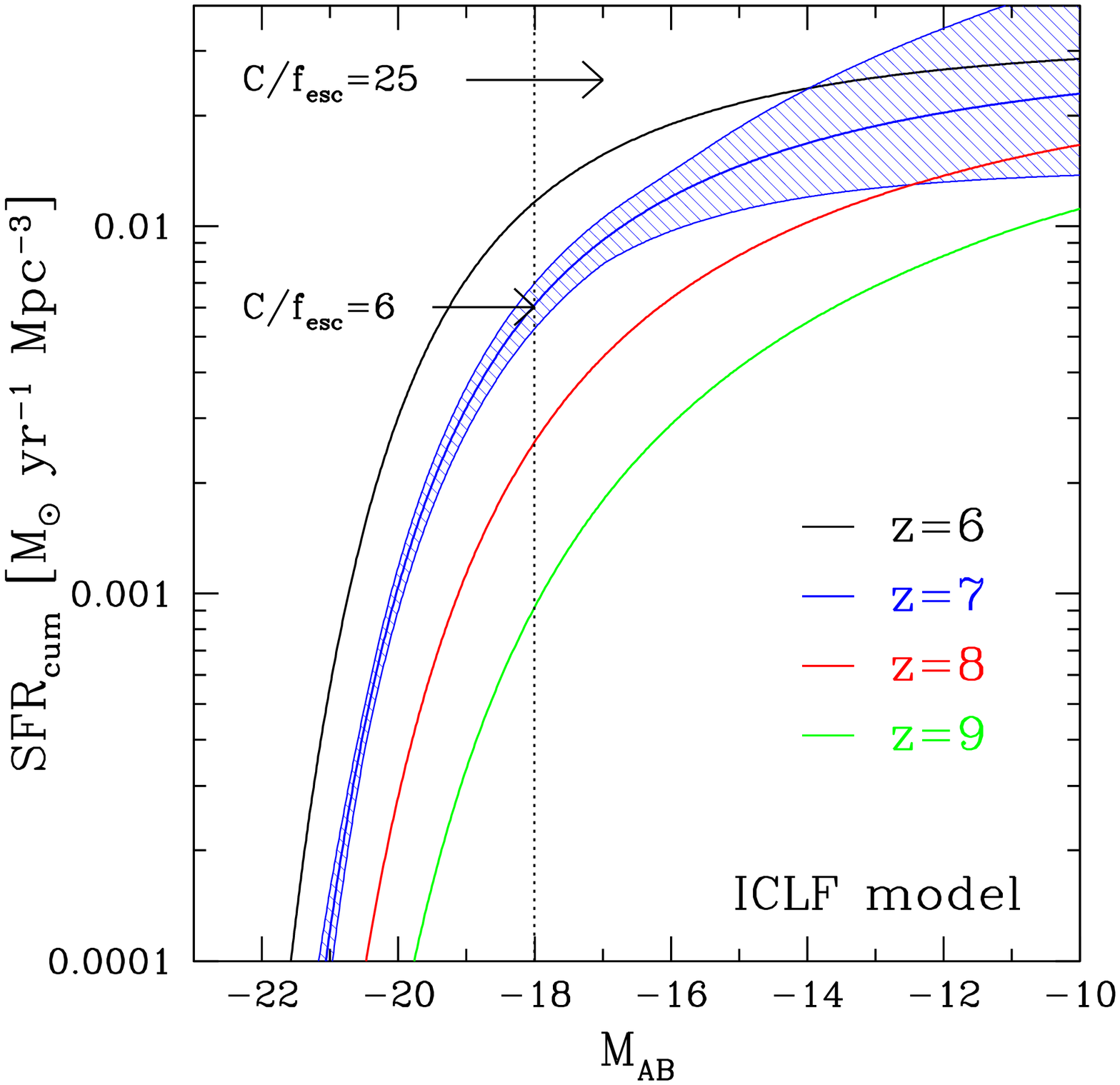}{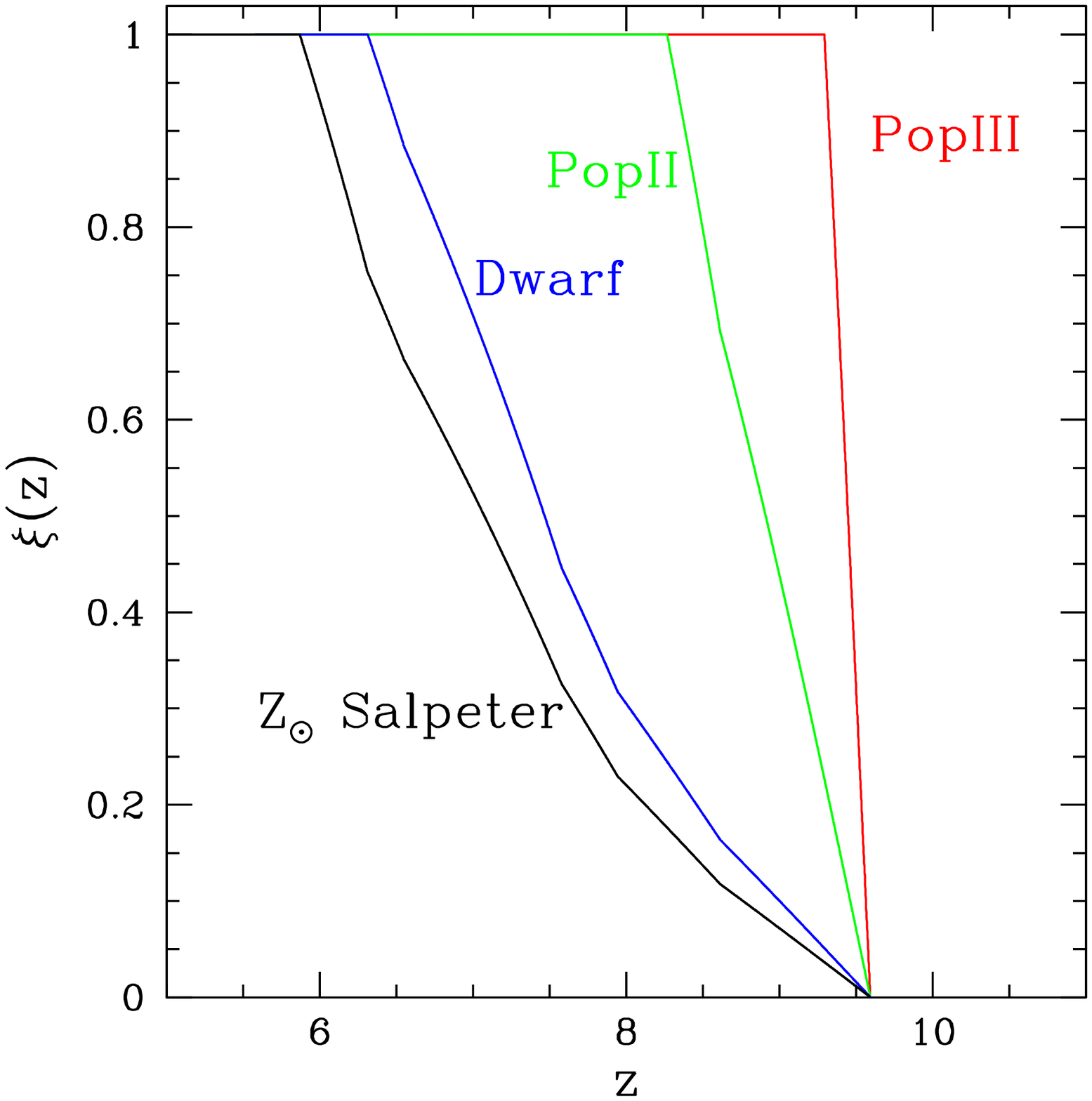} \caption{Left panel:
    Cumulative star formation rate for ICLF model with $\Delta t
    =200~\rm Myr$. Blue-shaded area gives $1\sigma$ uncertainty at $z=7$. Critical reionizing SFRs for $C/f_{esc}=6$ and
    $C/f_{esc}=25$ at $z=7$ are indicated.  Dotted line shows HUDF09 luminosity limit. Right panel: Evolution of
    IGM-mass reionization fraction $\xi(z)$  for
   LFs shown in left panel, integrated to $M_{AB}=-10$ under
   different SED assumptions (black: $Z=Z_{\sun}$, Salpeter; red:
   Pop III; blue: ``dwarf''-like metal-poor [SED CS1 of
   \citealt{schae03}]; green: Pop II,
   top-heavy; see \citealt{stiavelli04a}).
 }\label{fig:sfr}
\end{figure}

In this Letter we construct a model for the evolution of the galaxy
luminosity function at $z\gtrsim 4.5$ based on a modification of the
CLF method. We derive the relation between galaxy luminosity and
dark-matter halo mass at $z=6$, assuming a one-to-one correspondence
between observed galaxies and halos that formed in a period $\Delta t
= 200~\rm Myr$. Using $L(M_h)$ fixed at $z=6$, we derive the expected
LFs between $z=4$ and $z=9$, assuming only evolution of the underlying
dark-matter MF. The $z=5$ LF is consistent with observations, but our
model is less accurate at lower redshift because it underestimates the
faint-end slope. At $z \gtrsim 6$, we predict a moderate decrease of
$\phi_*$, a possible steepening of the faint-end slope, and continued
evolution of $L_*$ toward lower values (Figure~\ref{fig:lf} and
Table~\ref{tab:lf}). At all epochs, our predicted LF is well fitted by
a Schechter function with a prominent ``knee''.

The predicted number counts for the HUDF09-WFC3 field are a good match
to the dropouts observed at $z \sim 7$ and $z \sim 8$
(Table~\ref{tab:hudfcounts}). Overall our ICLF model is consistent
with the observed galaxy LF from $z \sim 5$ to $z \sim8$ with no
evolution in $L(M_h)$. DM halos assembly can explain LF evolution at
$z\geq 5$ without invoking a change in the properties of LBG star
formation. This is in agreement with the constant specific star
formation rate inferred at $z\gtrsim 5$ \citep{gonzalez09}.

Our model provides evidence for a reduced impact of feedback in
low-mass $z\gtrsim 5$ halos. In fact, we derive a star formation
efficiency weakly dependent on halo mass ($\eta \propto M_h^{0.3}$),
compared to the strong quenching of star formation derived at $z=0$
($\eta \propto M_h^3$; \citealt{cooray05a}), providing a testable
prediction for cosmological simulations. The strong suppression of
star formation in $M_h\lesssim 10^{11}~M_{\sun}$ halos suggested by
\citet{bouche10} and \citet{maiolino08} contrasts with $\alpha(z=6) \sim
1.7$ measured for halos with $M_h \gtrsim 2 \times 10^{10}~M_{\sun}$
(Section~\ref{sec:ml}). Such strong feedback would also imply that
galaxies appear incapable of sustaining reionization. In fact, with a
steep LF, sources below the HUDF-WFC3 detection limit may contribute
$\gtrsim 75\%$ of the ionizing flux, sufficient for full reionization
if $C/f_{esc} \lesssim 25$. A metal-poor and top-heavy IMF, or smaller
$C/f_{esc}$, are required to complete reionization at $z\gtrsim 8$ for
consistency with $(\tau_e)_{WMAP}$. While our extrapolation is
physically motivated to $T_{vir} \geqslant 2\times10^4~\mathrm{K}$,
it extends for $8$ magnitudes. Deeper observations are thus crucial to verify
that the LF faint-end remains steep.

\acknowledgements 

We thank the referee for useful suggestions. This work is supported by
grants NASA-ATP-NNX07AG77G, NSF-AST-0707474, STScI-HST-GO-11563,
STScI-HST-GO-11700, JWST-IDS-NAG5-12458 and by the Swiss National Science Foundation.




\clearpage
 
\begin{deluxetable}{rrrrrrrrrr}
\tablecolumns{10} 
\tablecaption{LF evolution\label{tab:lf}}
\tablehead{ \colhead{~} & \multicolumn{3}{l}{ {\small CLF model}}  & \multicolumn{3}{l}{ {\small  ICLF$_{200 \rm Myr}$ model} } &  \multicolumn{3}{l}{ {\small{ Observations}}} \\
\colhead{} &  \colhead{$(\phi_*)_{-3}$\tablenotemark{a}} & \colhead{$M_*$} & \colhead{$\alpha$}  &  \colhead{$(\phi_*)_{-3}$\tablenotemark{a}} & \colhead{$M_*$} & \colhead{$\alpha$} & \colhead{~~$(\phi_*)_{-3}$\tablenotemark{a}} & \colhead{$M_*$} & \colhead{$\alpha$}  }
\startdata 
\sidehead{Input LF}
$z=6$ & $1.4$ & $-20.24$ & -1.74 & $1.4$ & $-20.24$ & -1.74 & {\small $1.4 \pm 0.5$ }& {\small $-20.24 \pm 0.19$} & {\small $-1.74 \pm 0.16$}\\  
\sidehead{Predicted LF}
$z=4$ & $3.4 $& $-20.81$ & -1.60  & $1.3$ & $-20.90$ & -1.57 & {\small $1.3 \pm 0.2$ }& {\small $-20.98 \pm 0.10$} & {\small $-1.73 \pm 0.05$}\\  
$z=5$ & $2.3$ & $-20.51$  & -1.66  & $1.5$ & $-20.55$ & -1.63 & {\small $1.0 \pm 0.3$} & {\small $-20.64 \pm 0.13$ }&  {\small $-1.66 \pm 0.09$}\\  
$z=7$ & $0.69$ &$-20.00$ & -1.84  & $1.0$ &$-20.00$ & -1.84 & & &\\  
$z=8$ & $0.36$& $-19.75$ & -1.89 & $0.60$& $-19.70$ & -1.90 & & &\\
$z=9$ & $0.14$& $-19.50$ & -1.99  & $0.22$& $-19.55$ & -2.22\tablenotemark{b} & & & \\
\enddata
\tablecomments{Best-fit Schechter parameters for LFs
  in Figure~\ref{fig:lf} (second to fourth column: CLF model;
  fifth to seventh column: ICLF; last three columns: \citealt{bouwens07} measurements). Fit holds for $-22.5 \leqslant M_{AB} \leqslant - 18$. Relative residuals are $\lesssim 20\%$.}
\tablenotetext{a}{Units: $10^{-3}~\mathrm{Mpc^3}$ }
\tablenotetext{b}{Asymptotic faint-end slope is $\alpha \sim -2$ }
\end{deluxetable}

\begin{table} \begin{center} \caption{Predicted dropouts for HUDF09 field\label{tab:hudfcounts}}
\smallskip
\smallskip
\begin{tabular}{cccccc}
\tableline\tableline
& { Observed } & { $\mathrm{ICLF_{200Myr}}$}&  $\mathrm{ICLF_{100Myr}}$ &  $\mathrm{ICLF_{300Myr}}$ &  CLF \\
\tableline
z-drop & { 16 } & $13.4 \pm 5.8$ & $16.6 \pm 6.8$  & $11.3 \pm 5.2$ & $9.8 \pm 4.7$  \\  
Y-drop & {  5 } & $5.3 \pm 3.1$ & $8.5 \pm 4.3$ & $3.9 \pm 2.5$ & $3.2 \pm 2.7$ \\
\tableline \tableline
\end{tabular}
\tablecomments{Observed \citep{oesch09_zdrop,bouwens09_ydrop} and
  predicted number counts for galaxies at
  $z\sim 7$ (z-dropouts) and $z\sim 8$ (Y-dropouts) in the HUDF09
  field for different
  (I)CLF models. Predictions include convolution with effective
  HUDF09 volume as function of source magnitude.
  $1\sigma$ uncertainty includes cosmic variance.}
\end{center}
\end{table}


\begin{thebibliography}{43}
\expandafter\ifx\csname natexlab\endcsname\relax\def\natexlab#1{#1}\fi

\bibitem[Bouch\'e et al.(2010)]{bouche10} Bouch\'e N. et al. 2010, \apj, submitted, arXiv:0912.1858

\bibitem[{{Bolton} \& {Haehnelt}(2007)}]{bolton07}
{Bolton}, J.~S. \& {Haehnelt}, M.~G. 2007, \mnras, 382, 325

\bibitem[{Bouwens} {et~al.}(2007)]{bouwens07}
{Bouwens}, R.~J. at al. 2007, \apj, 670, 928

\bibitem[{{Bouwens} {et~al.}(2008)}]{bouwens08}
---. 2008, \apj, 686, 230

\bibitem[{{Bouwens} {et~al.}(2010{\natexlab{a}}){Bouwens}, {Illingworth},
  {Oesch}, {Stiavelli}, {van Dokkum}, {Trenti}, {Magee}, {Labbe}, {Franx}, \&
  {Carollo}}]{bouwens09_ydrop}
{Bouwens}, R.~J. et al. 2010{\natexlab{a}}, \apjl, 709, 133

\bibitem[{{Bouwens} {et~al.}(2010{\natexlab{b}}){Bouwens}, {Illingworth},
  {Oesch}, {Trenti}, {Stiavelli}, {Carollo}, {Franx}, {van Dokkum}, {Labbe}, \&
  {Magee}}]{bouwens09_slope}
{Bouwens}, R.~J. et al. 2010{\natexlab{b}}, \apjl, 708, 69

\bibitem[{{Bouwens} {et~al.}(2004){Bouwens}, {Illingworth}, {Thompson},
  {Blakeslee}, {Dickinson}, {Broadhurst}, {Eisenstein}, {Fan}, {Franx},
  {Meurer}, \& {van Dokkum}}]{bouwens04}
{Bouwens}, R.~J. et al. 2004, \apjl, 606, L25

\bibitem[{{Bunker} {et~al.}(2009){Bunker}, {Wilkins}, {Ellis}, {Stark},
  {Lorenzoni}, {Chiu}, {Lacy}, {Jarvis}, \& {Hickey}}]{bunker09}
{Bunker}, A. et al. 2009, ArXiv:0909.2255

\bibitem[{{Bunker} {et~al.}(2004){Bunker}, {Stanway}, {Ellis}, \&
  {McMahon}}]{bunker04}
{Bunker}, A.~J. et al. 2004, \mnras, 355, 374

\bibitem[Cantalupo(2010)]{cantalupo10} Cantalupo, S. 2010, \mnras, in press, 

\bibitem[{{Cooray}(2005)}]{cooray05b}
{Cooray}, A. 2005, \mnras, 364, 303

\bibitem[{{Cooray} \& {Milosavljevi{\'c}}(2005)}]{cooray05a}
{Cooray}, A. \& {Milosavljevi{\'c}}, M. 2005, \apjl, 627, L89

\bibitem[{{Cooray} \& {Ouchi}(2006)}]{cooray06}
{Cooray}, A. \& {Ouchi}, M. 2006, \mnras, 369, 1869

\bibitem[Dav\'e et al.(2008)]{dave08} Dav\'e, R. et al. 2008, \mnras, 391, 110


\bibitem[Finkelstein et al.(2009)]{fink09} {Finkelstein}, S.~L. at al. 2009, arXiv0912.1338


\bibitem[{{Gonzalez} {et~al.}(2010){Gonzalez}, {Labbe}, {Bouwens},
  {Illingworth}, {Franx}, {Kriek}, \& {Brammer}}]{gonzalez09}
{Gonzalez}, V. et al. 2010, \apj, in press, arXiv:0909.3517

\bibitem[{{Haardt} \& {Madau}(1996)}]{haardt96}
{Haardt}, F. \& {Madau}, P. 1996, \apj, 461, 20

\bibitem[{{Hopkins} \& {Elvis}(2009)}]{hopkins09}
{Hopkins}, P.~F. \& {Elvis}, M. 2009, \mnras, 1516

\bibitem[Kere{\v s} et al.(2005)]{keres05} Kere{\v s}, D. et al.
      2005, \mnras, 363, 2

\bibitem[{{Komatsu} {et~al.}(2009){Komatsu}, {Dunkley}, {Nolta}, {Bennett},
  {Gold}, {Hinshaw}, {Jarosik}, {Larson}, {Limon}, {Page}, {Spergel},
  {Halpern}, {Hill}, {Kogut}, {Meyer}, {Tucker}, {Weiland}, {Wollack}, \&
  {Wright}}]{komatsu09} {Komatsu}, E. et al. 2009, \apjs, 180, 330

\bibitem[{{Labb{\'e}} {et~al.}(2009){Labb{\'e}}, {Gonzalez}, {Bouwens}, {Illingworth},
  {Oesch}, {van Dokkum}, {Carollo}, {Franx}, {Stiavelli}, {Trenti}, {Magee}, \&
 {Kriek}}]{labbe09a}
 {Labb{\'e}}, I. et al. 2009, ArXiv:0910.0838

\bibitem[Lee et al.(2009)]{lee09} Lee, K.~S. et al. 2009, \apj, 695, 368 

\bibitem[{{Madau} {et~al.}(1999){Madau}, {Haardt}, \& {Rees}}]{madau99}
{Madau}, P. et al. 1999, \apj, 514, 648

\bibitem[{{Madau} {et~al.}(1998){Madau}, {Pozzetti}, \& {Dickinson}}]{madau98}
{Madau}, P. et al. 1998, \apj, 498, 106

\bibitem[{{Madau} {et~al.}(2004){Madau}, {Rees}, {Volonteri}, {Haardt}, \&
  {Oh}}]{madau04}
{Madau}, P. et al. 2004, \apj, 604, 484

\bibitem[Maiolino et al.(2008)]{maiolino08} Maiolino et al. 2008, \aap, 488, 463

\bibitem[{{McLure} {et~al.}(2010){McLure}, {Dunlop}, {Cirasuolo}, {Koekemoer},
  {Sabbi}, {Stark}, {Targett}, \& {Ellis}}]{mclure09}
{McLure}, R.~J. et al. 2010, \mnras, in press, arXiv:0909.2437


\bibitem[{{Oesch} {et~al.}(2010{\natexlab{a}}){Oesch}, {Bouwens}, {Carollo},
  {Illingworth}, {Trenti}, {Stiavelli}, {Magee}, {Labbe}, \&
  {Franx}}]{oesch09_size}
{Oesch}, P.~A. et al. 2010{\natexlab{a}}, \apjl, 709, 21

\bibitem[{{Oesch} {et~al.}(2010{\natexlab{b}}){Oesch}, {Bouwens},
  {Illingworth}, {Carollo}, {Franx}, {Labbe}, {Magee}, {Stiavelli}, {Trenti},
  \& {van Dokkum}}]{oesch09_zdrop}
{Oesch}, P.~A. et al. 2010{\natexlab{b}}, \apjl, 709, 16

\bibitem[{{Oesch} {et~al.}(2009{\natexlab{c}}){Oesch}, {Carollo}, {Stiavelli},
  \& {Trenti}}]{oesch09}
{Oesch}, P.~A. et al. 2009{\natexlab{c}}, \apj, 690, 1350

\bibitem[{{Overzier} {et~al.}(2006){Overzier}, {Bouwens}, {Illingworth}, \&
 {Franx}}]{overzier06}
{Overzier}, R.~A. et al. 2006, \apjl, 648, L5

\bibitem[{{Pawlik} {et~al.}(2009){Pawlik}, {Schaye}, \& {van
  Scherpenzeel}}]{pawlik09}
{Pawlik}, A.~H. et al. 2009, \mnras, 394,
  1812

\bibitem[Ricotti et al.(2008)]{ricotti08} Ricotti M. et al. (2008), \apj, 685, 21

\bibitem[{{Salpeter}(1955)}]{salpeter}
{Salpeter}, E.~E. 1955, \apj, 121, 161

\bibitem[Schaerer et al.(2003)]{schae03} Schaerer, D. et al. 2003,
  \aap, 397, 527

\bibitem[{{Sheth} \& {Tormen}(1999)}]{st99}
{Sheth}, R.~K. \& {Tormen}, G. 1999, \mnras, 308, 119

\bibitem[{{Shull} \& {Venkatesan}(2008)}]{shull08}
{Shull}, J.~M. \& {Venkatesan}, A. 2008, \apj, 685, 1

\bibitem[{{Sokasian} {et~al.}(2004){Sokasian}, {Yoshida}, {Abel}, {Hernquist},
  \& {Springel}}]{sokasian04}
{Sokasian}, A., et al.  2004, \mnras, 350, 47

\bibitem[Stark et al.(2007)]{stark07} Stark, D.~P. et al. 2007, \apj, 668, 627

\bibitem[{{Stark} {et~al.}(2009){Stark}, {Ellis}, {Bunker}, {Bundy}, {Targett},
  {Benson}, \& {Lacy}}]{stark09}
{Stark}, D.~P. et al. 2009, \apj, 697, 1493

\bibitem[{{Stiavelli} {et~al.}(2004{\natexlab{a}}){Stiavelli}, {Fall}, \&
  {Panagia}}]{stiavelli04a}
{Stiavelli}, M. et al. 2004{\natexlab{a}}, \apj, 600,
  508

\bibitem[{{Stiavelli} {et~al.}(2004{\natexlab{b}}){Stiavelli}, {Fall}, \&
  {Panagia}}]{stiavelli04b}
---. 2004{\natexlab{b}}, \apjl, 610, L1

\bibitem[{{Tegmark} {et~al.}(1997){Tegmark}, {Silk}, {Rees}, {Blanchard},
  {Abel}, \& {Palla}}]{tegmark97}
{Tegmark}, M. et al. 1997, \apj, 474, 1

\bibitem[{{Trenti} \& {Stiavelli}(2008)}]{trenti07}
{Trenti}, M. \& {Stiavelli}, M. 2007, \apj, 667, 38

\bibitem[{{Trenti} \& {Stiavelli}(2008)}]{trenti08}
---. 2008, \apj, 676, 767

\bibitem[{{Trenti} \& {Stiavelli}(2009)}]{ts09}
---. 2009, \apj, 694, 879

\bibitem[Trenti et~al.(2008)]{tss08}
{Trenti}, M. et al. 2008, \apj, 687, 1 

\bibitem[{{Trenti} {et~al.}(2009){Trenti}, {Stiavelli}, \& {Shull}}]{trenti09b} {Trenti}, M. et al. 2009, \apj, 700, 1672

\bibitem[{{Vale} \& {Ostriker}(2004)}]{vale04}
{Vale}, A. \& {Ostriker}, J.~P. 2004, \mnras, 353, 189

\bibitem[{{Wechsler} {et~al.}(2001){Wechsler}, {Somerville}, {Bullock},
  {Kolatt}, {Primack}, {Blumenthal}, \& {Dekel}}]{wechser01}
{Wechsler}, R.~H. et al. 2001, \apj, 554, 85

\end{thebibliography}
\end{document}